# SoundLoc: Acoustic Method for Indoor Localization without Infrastructure

Ruoxi Jia
University of California, Berkeley
Berkeley, CA 94720
ruoxijia@berkeley.edu

Ming Jin
University of California, Berkeley
Berkeley, CA 94720
jinming@berkeley.edu

Costas J. Spanos
University of California, Berkeley
Berkeley, CA 94720
spanos@berkeley.edu

## ABSTRACT

Identifying locations of occupants is beneficial to energy management in buildings. A key observation in indoor environment is that distinct functional areas are typically controlled by separate HVAC and lighting systems and room level localization is sufficient to provide a powerful tool for energy usage reduction by occupancy-based actuation of the building facilities. Based upon this observation, this paper focuses on identifying the room where a person or a mobile device is physically present. Existing room localization methods, however, require special infrastructure to annotate rooms.

SoundLoc is a room-level localization system that exploits the intrinsic acoustic properties of individual rooms and obviates the needs for infrastructures. As we show in the study, rooms' acoustic properties can be characterized by Room Impulse Response (RIR). Nevertheless, obtaining precise RIRs is a time-consuming and expensive process. The main contributions of our work are the following. First, a cost-effective RIR measurement system is implemented and the Noise Adaptive Extraction of Reverberation (NAER) algorithm is developed to estimate room acoustic parameters in noisy conditions. Second, a comprehensive physical and statistical analysis of features extracted from RIRs is performed. Also, SoundLoc is evaluated using the dataset consisting of ten (10) different rooms. The overall accuracy of 97.8% achieved demonstrates the potential to be integrated into automatic mapping of building space.

## Categories and Subject Descriptors

H.5.5 [**Information Interfaces and Presentation**]: Sound and Music Computing – *Signal analysis, synthesis, and processing.*
H.4.1 [**Information Systems Applications**]: Office Automation

## 1. INTRODUCTION

Commercial buildings contribute to 19% of the primary energy consumption in US. Prior research has shown that most of buildings use static control for building facilities, such as Heating, Ventilation and Air Conditioning (HVAC) and lighting systems, thereby considerable energy is wasted in conditioning and lighting unoccupied spaces [1, 2]. Awareness of occupancy information can help adaptively run the conditioning and lighting systems and reduce energy consumption in buildings. Therefore, availability of indoor locations has become an immediate need.

Unlike outdoors, where GPS can provide a relatively accurate and robust solution for positioning, indoor localization has not been equally facilitated by GPS due to significant positioning error of satellite based navigation systems in closed environments. A variety of alternatives have been proposed for indoor operation, ranging from visual [3] to infrared [4]. There have also been extensive research works focusing on indoor localization systems based on WiFi wireless network along with WiFi enabled devices [5]. However, the density of access points has a strong influence on localization accuracy. The reported WiFi localization accuracy drops below 70% in real usage environments since access point density may be low or occupancy variations may lead to significant WiFi signal variations. Also, these techniques face certain disadvantages that special-purpose infrastructure is required.

In contrast to infrastructure-based techniques, SoundLoc is supported by internal microphone and speaker on laptops or mobile phones, which are the most ubiquitous devices. A key observation that supports our work is that indoor environment is well-structured and can be organized into areas with distinct geometry and functionalities. These areas can either be open spaces without a proper boundary like hallway, or closed spaces such as offices. "Room" refers to a closed space in most cases. We notice that the control of lighting and HVAC systems are typically room-level based. Therefore, instead of notating location with physical coordinates (latitude/longitude), a room level localization is sufficient for any occupancy-based control of lighting and HVAC systems. Radio-based techniques have birth defects of confusing nearby rooms as the inference for location is based on received WiFi signal strength, which varies with time and indoor environment changes. To overcome this shortcoming, our work exploits the acoustic properties to identify a location. The acoustic effects of a space are governed by geometry and furnishings. Even though two rooms are geospatially adjacent, they can be easily distinguishable in acoustic feature space.

This article describes SoundLoc, a room identification method based on the extraction of acoustic features of rooms. The concept of "room" also incorporates open spaces, which exhibit special acoustic properties that can be leveraged for localization as well. SoundLoc exploits the acoustic effects of a room on audio signals and can be freely implemented without any infrastructures.

The rest of this paper is organized as follows. Section 2 describes related work. Section 3 describes the formulation of localization problems in terms of rooms' acoustic properties and explores various acoustic features that are promising to be used for localization. Section 4 describes the experiment design. Section 5 evaluates the performance of SoundLoc. Section 6 concludes the paper.

## 2. RELATED WORK

Indoor localization has been extensively studied. Early works tried to build an empirical RF propagation model to estimate the location by received signal strength from multiple known access points [6]. This method suffers from meters of localization error and the model is very complex in order to take the dynamics of RF environment into account. Recent works focus on giving a quantitative description of the indoor environment, i.e. to create a unique fingerprint for a given room. Various types of fingerprints have been developed, mainly RF fingerprints and ambient fingerprints.

Radar [6] pioneered fingerprinting method based on received signal strength hearing from multiple access points. Horus [7] developed a more sophisticated and accurate approach wherein the signal strength measurements at each location are represented as a probability distribution and fingerprints are matched using maximum likelihood criterion instead of a deterministic matching in Radar. ARIEL [8] proposed a room localization system correlate occupants' motion pattern with WiFi signal to improve identification accuracy. However, these techniques are hampered by long-term signal variations caused by occupancy variations and low access point density. SurroundSense [9] extends the fingerprinting idea by combining multiple sensors on smartphone and building a map using several ambient features such as background sound, light, color in addition to WiFi signal strength. In SurroundSense, the acoustic fingerprint is a compact histogram of sound in the time domain. ABS [10] alternatively exploits the spectral representation of background sounds for room identification. SoundSense [11] also uses spectral features but transient sounds are used instead to classify sounds observed on a mobile device. However, these works are based upon the assumption that the ambient sound features in a place can be stationary and suggestive, which is not always satisfied in applications. In real cases, the background sound can vary both transiently and long-termly. People's talking could appear randomly and even different HVAC on/off states could generate distinctive background sound that has influence on the ambient sound fingerprints. Therefore, sounds fingerprints are often combined with other localization techniques in order to achieve a higher accuracy. ABS [10] integrates background sound fingerprints with WiFi and 69% has been realized. Several efforts have been made to identify the location by using pre-installed equipment to detect the sound generated by occupants or mobile devices. Daredevil [12] deploys microphone arrays to determine the location of smartphones emitting a tone by triangulation.

Rather than detecting the uncontrolled background sound, acoustic fingerprint exploits the room's acoustical effect on audio signals. [13] demonstrates a system that identifies the room through analysis of acoustical properties of audio recordings. 61% accuracy has been achieved for musical signals and 85% for speech signals. [14] classifies the room based on reverberation time extracted from RIRs, and 3.9 % error rate has been achieved. However, the RIR samples used in the paper are collected from places that vary significantly in volumes and inside furnishings, such as classrooms, music hall, auditorium etc.

Our work differs from the above work in following aspects. Firstly, no extra microphones need to be installed as we employ internal microphone and speaker on the device. Secondly, we leverage the rooms' intrinsic acoustic properties rather than analyzing the non-stationary background. Thirdly, we present a real-time and cost-effective RIR measurement system. Instead of using RIR samples available online that were collected from places varying considerably in volumes, we perform RIR measurement in several similar building environments such as adjacent offices and demonstrate the potential of acoustic features to identify places in buildings, both closed and open spaces are included.

## 3. THEORY FRAMEWORK

Our localization scheme is based on the acoustic features extracted from the Room Impulse Response (RIR), which is an important characterization of the acoustical effect of a room on audio signals. RIR is related to the room's size, shape, surface absorption, etc. However, obtaining accurate RIRs is a time-consuming process and requires special measurement equipment, such as dodecahedron loudspeaker and soundfield microphone. In this paper, we acquire RIRs by using built-in speakers and microphones on laptops. Not surprisingly, the RIRs we measured are very noisy, which can distort the evaluated acoustic parameters. In this section, we present a variety of acoustic features extracted from the RIR and their relation to rooms' geometric and absorption properties. Also, we put forward a novel noise compensation technique to extract features from noisy RIRs.

### 3.1 Problem formulation

When a sound is produced inside a room, the sound signal travels not only the direct path from source to receiver, but also arrives at the receiver after several bouncing off walls or other objects. Therefore, the signal received is a superposition of multiple delay and distorted versions of the original signal, which is perceived as echo and reverberation. Intuitively, the received signal contains information about room's size and absorption properties. Since the environment's geometry and interior furnishing materials are roughly linear and time-invariant, the "room effect" can be viewed as a linear time-invariant system and characterized by an impulse response $h(t)$. Thus, the received signal is a convolution of the transmitted signal and a room impulse response (RIR) in the time domain, as illustrated in Fig. 1. Since there exists a one-to-one mapping from a room to its "room effect", a unique label can be assigned to it theoretically if its RIR is available. In other words, we can work out the indoor localization problem with a granularity of rooms if the RIR of a unknown location is obtained.

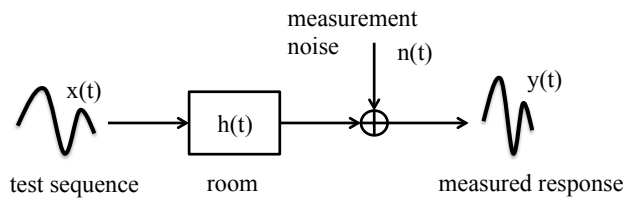

**Figure 1.** The room can be modeled as a linear time-invariant system and the received signal is a linear convolution of test excitation sequence and room impulse response.

A common approach to measure the RIR is to apply a known excitation signal, measure the system's output and then deconvolve the measured response with the test sequence. The choice concerning the excitation signals and deconvolution techniques is of essential significance to RIR measurement. Several types of most commonly used excitation signals are presented and compared in [15]. In our paper, we utilize Maximum Length Sequence (MLS) as the input sequence, which is known for its capability of providing vastly superior dynamic range and high signal-to-noise ratio. MLS is a periodic pseudo-random signal and behaves almost the same as white noise in the sense of schochastic properties. Hence we can acquire the RIR by computing the autocorrelation of the received signal [16],

$$h(k) = R_y(k) = E\left[y(n)y^*(n-k)\right] \quad (1)$$

where $R_y(k)$ denotes the autocorrelation and it is defined as the expectation of the product of the original signal and a delay of the signal's complex conjugate. In order to reduce the time-aliasing error, a MLS with longer period is preferred [17]. In our measurement system, the length of MLS is $2^{17}-1$. And we perform the autocorrelation using the fast Hadamard transform in order to minimize the computing time and efficiently determine the RIRs [18].

Hereto, a theoretical framework for RIR acquisition is introduced. However, RIRs take the form of time series and cannot be directly fed in the classification algorithm. It is necessary to extract some "valuable" features from RIR and these features should include rich information regarding location. And furthermore an acoustic fingerprint can be built by combining these acoustic features and used to uniquely determine the room label.

**Table 1. Details of investigated rooms**

| Investigated Area | Area (m²) | Description |
|---|---|---|
| Office A | 10.2 | Closed space, quiet |
| Office B | 8.8 | Closed space, quiet |
| Office C | 7.1 | Closed space, quiet |
| Office D | 11.7 | Closed space, quiet |
| Conference Room | 26.1 | Closed space, quiet |
| Lab | 11.7 | Closed space, server noise |
| Kitchen | 6.5 | Open space, speaking and coffee machine noise |
| Hallway | - | Open space, speaking noise |
| Stairs | Significantly larger than other closed spaces | Closed space, speaking, footstep and door creaky noise |
| Cubicle Zone | 8.8 | Open space surrounded by clipboard, speaking noise |

## 3.2 Acoustic feature exploration

### 3.2.1 Temporal features

We use kurtosis of the RIR in time domain to remark its temporal properties. In statistics, kurtosis describes the peakedness of the probability density function of a real-valued random variable. Kurtosis of a signal $x(k)$ in the time domain is defined as follows:

$$kur[x(k)] = \frac{E[x(k)-\mu]}{\sigma^4} \quad (2)$$

where $E[]$ is the expectation operator, $\mu$ is the mean of signal and $\sigma$ is the standard deviation. Higher kurtosis of a time domain signal means more of the variance results from infrequent extreme deviations, in contrast to frequent modestly sized deviations. The kurtosis of the RIR is an indicator of the volume of a room. If the room is large, then the RIR will have infrequent large deviations and thereby higher kurtosis. On the contrary, a small volume will result in lower kurtosis. Fig. 2 illustrates the distributions of temporal kurtosis in different locations. The details of locations involved are listed in Table 1. As can be seen, closed spaces with relatively small volumes exhibit small kurtosis in the time domain, while open spaces or spaces with large volume have larger kurtosis.

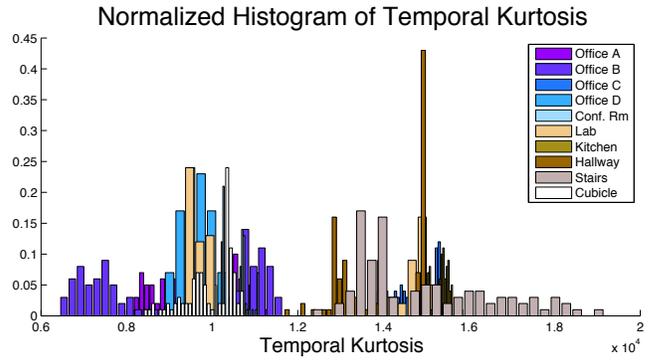

**Figure 2.** Normalized histogram of temporal kurtosis. Closed spaces with small volumes such as offices, conference room, and cubicle have smaller temporal kurtosis. Open spaces such as kitchen and hallway or closed spaces with large volume such as stairs have large kurtosis.

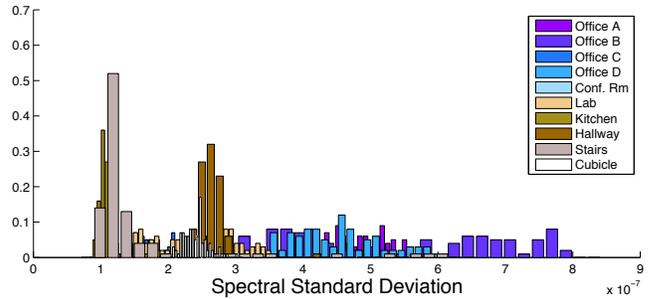

**Figure 3.** Normalized histogram of spectral standard deviation in octave band centered at 250 Hz. Places covered with carpet such as offices, lab, hallway and cubicle exhibit a larger spectral standard deviation. Places without special sound reduction such as stairs and kitchen shows a smaller spectral standard deviation.

### 3.2.2 Spectral features

In audio physics, direct-to-reverberant energy ratio is an important parameter to characterize a room's acoustic properties such as diffuseness. It depends on the geometry and absorption of the space where the sound waves propagate. In [19], it is shown that the direct-to-reverberant energy ratio increases with the standard deviation of RIR spectrum. Hence, spectral standard deviation can be used to characterize a room. Since the absorption properties of materials are a function of frequency, we further inspect this feature within different frequency bands, as defined by

$$std_{[f_1,f_2]}[H(f)] = E_{[f_1,f_2]}[H^2(f)] - E^2_{[f_1,f_2]}[H(f)] \quad (3)$$

where $H(f)$ denotes the Fourier transform of the RIR and $E_{[f_1,f_2]}[]$ means taking the mean value over the frequency band ranging from $f_1$ to $f_2$. The distributions of spectral standard deviation in different rooms are plotted in Fig. 3. The places investigated here exhibits different absorption properties. For instance, offices, lab, cubicle, conference room and hallway are covered with carpet, which is a good sound-absorbent material. In these areas, sound energy is absorbed before it can bounce off in the space and generates reverberation. Direct sound energy from the emitter to the receiver will dominate in this case, and thereby the locations above have a higher direct-to-reverberant sound ratio, i.e. a larger spectral standard deviation. In contrast, locations without special sound reduction, such as stairs and kitchen, show a relatively small spectral standard deviation.

In addition, the kurtosis of Fourier coefficients is also included in our feature pool. Since the room can be identified by its room modes, which are collection of resonances that exist in a room when it is excited by a sound source. Room modes can be noticed by magnitude peaks in the spectrum of the RIR. We use the kurtosis of Fourier coefficients to describe room modes and further to characterize a room.

### 3.2.3 Energetic features

In general, energetic features of RIRs describe how sound energy decays as it propagates in rooms. Reverberation time (RT) is a promising energetic feature for room identification as it doesn't require a special microphone arrangement nor does it rely on the source orientation [20]. A standard RT is defined by the time taken for the acoustic energy in a space to decay by 60 dB once the source is turned off. According to Sabine's formula [21],

$$RT = 0.161 \frac{V}{S\alpha} \quad (4)$$

RT is directly related to the volume $V$ and the surface absorption of the room, given by the product of surface area $S$ and average absorption coefficients $\alpha$. RT can be estimated from the normalized energy decay curve (EDT), which is computed by reverse integrating the squared RIR,

$$E(t) = G \int_t^\infty h^2(\tau) d\tau \quad (5)$$

where $G$ is a constant related to excitation level. Then, RT can be given by estimating the decay rate over $[-5dB, -35dB]$ using linear regression techniques. ISO 3382 specifies the above measurement method as a standard. However, this method is not applicable in our case. Firstly, the RIR we collect is very noisy. The noise stems both from the measurement equipment and the background. Noise dominates and stretches the energy decay curve as shown in Fig. 4. Secondly, the positions of the speaker and the microphone are very close to each other on laptops, which results in a very strong direct feed-through. It shows as a sharp drop at the beginning of both RIR and EDT. However, this segment makes no contribution to calculation of reverberation time, since the direct sound energy depends only on the distance between speaker and microphone and is independent of rooms' properties. The EDT segment that is useful for RT calculation is where reverberation dominates, but it is very short as illustrated in Fig. 4. The decay of sound energy is less than -10 dB in this segment. Therefore, the traditional method does not work in our case.

ISO 3382 international standard also defines three different methods to compensate the noise effects for the calculation of the EDT. It has been shown that these methods are ill-suited when the peak-signal to noise ratio (PNR) is smaller than 45 dB [22]. In our case, the average PNR of RIR samples is 34 dB. Therefore, a new noise compensation method is needed to extract reverberation time from very noisy RIRs.

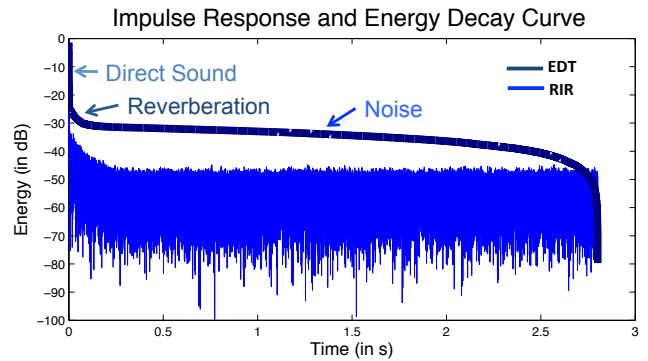

**Figure 4.** Impulse response and energy decay curve. EDT segments corresponding to direct sound domination, reverberation domination, noise domination are annotated.

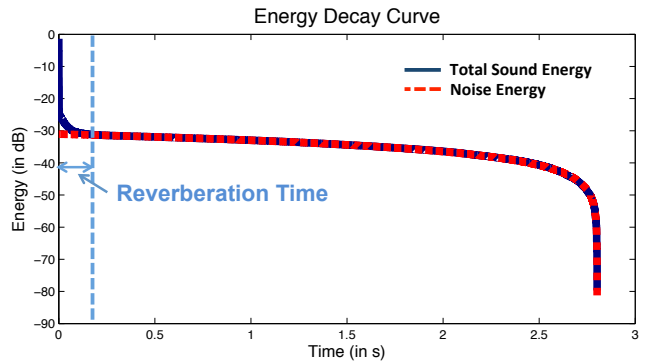

**Figure 5.** NAER estimate the noise energy and defines RT as the time taken for total sound energy decay to noise energy

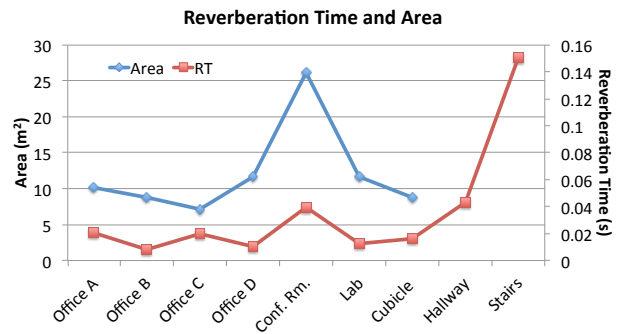

**Figure 6.** RT given by NAER varies in the same trend as the area values. Particularly, we do not have access to precise area values of hallway and stairs, but they are much larger than other locations investigated and these two locations also have much larger RTs as is shown here.

To overcome the aforementioned obstacles, we introduce the Noise Adaptive Extraction of Reverberation (NAER) method. NAER gives an estimation of noise level and defines reverberation time as the time taken for sound energy decays to noise level (Fig. 5). Therefore, it is robust to noise level of the environment. The algorithm of NAER is presented in Table 2. Based upon NAER, we calculate reverberation time of several rooms in listed in Table. 1, the result shows a consistence with the volume and absorption properties of rooms, as illustrated in Fig.6.

**Table 2. Pseudocode for NARE**

| |
|---|
| NARE_Measurement (*RIR, PerNoise, BondP, Th*) |
| **Inputs**:  *RIR*: room impulse response of length L |
|  *PerNoise*: the last PerNoise portion of RIR to estimate noise level |
|  *BondP*: the bonding point defined where sound energy meets noise |
|  *Th*: threshold to define reverberation time |
| <u>Noise Estimation:</u> |
|  *NoiseLevel* ← *RIR* (*PerNoise : end*) |
| <u>Pseudo Noise Energy Curve Calculation:</u> |
|  *PseudoNoiseEnergy* ← inverse integrate *NoiseLevel* |
|  *SoundEnergy* ← inverse integrate *RIR* |
|  *PseudoNoiseEnergy* ← *PseudoNoiseEnergy* + *SoundEnergy(BondP)* − *PseudoNoiseEnergy(BondP)* |
| <u>Reverberation Time Extraction:</u> |
| **for** *FindInd* ∈ {1,...L} |
| **if** *SoundEnergy(FindInd)* − *PseudoNoiseEnergy(FindInd)* < *Th* |
|  **break** |
| **end of for** |

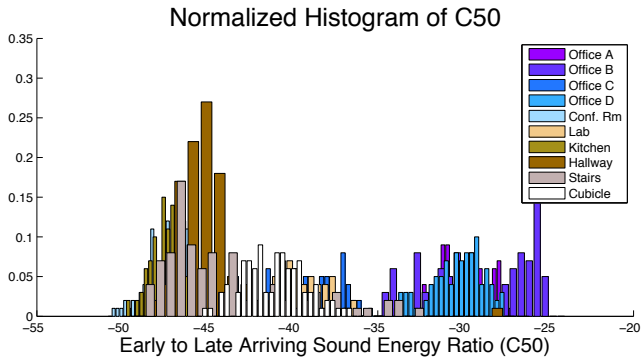

**Figure 7.** Normalized histogram of early to late arriving sound energy ratio. Sound energy in rooms with smaller volumes tends to be dominated by early energy, while C50 for closed spaces with larger volumes or open spaces tends to be smaller.

Chu [23] proposed to subtract noise from the squared RIR before backwards integration, which makes the EDC ascend at some time slots if the noise level is high. Comparatively speaking, NAER doesn't distort the energy decay curve and is superior when RIRs are noisy and have a very strong direct sound energy. This condition can be viewed on EDC by a narrow reverberation dominated zone that is squeezed between a sheer direct sound energy drop zone and a gentle noisy energy decay zone.

We also extract RT in various octave bands in order to take into account the frequency-dependent absorption properties of the room. In addition, early to total sound energy ratio (D50), early to late arriving sound energy ratio (C50) and center time of the squared impulse response (TS) are also used for energetic features. Generally speaking, these parameters describe where the sound energy is concentrated along the timeline. The dominance of early energy is an indicator for a smaller volume or a low sound absorption (Fig. 7). D50, C50, TS can be computed by the following formulas, respectively:

$$D_{50} = \frac{\int_0^{0.05s} h^2(t)dt}{\int_0^\infty h^2(t)dt} \qquad (6)$$

$$C_{50} = 10\log\left(\frac{D_{50}}{1-D_{50}}\right) \qquad (7)$$

$$TS = \frac{\int_0^\infty t \cdot h^2(t)dt}{\int_0^\infty h^2(t)dt} \qquad (8)$$

## 4. EXPERIMENTS

The aim of our experiments is to verify if the noisy RIR's obtained by the cheap internal speakers and microphones on laptops contain "valuable" features that are capable of indicating indoor locations. Furthermore, we design experiment to test the noise-robustness and time-invariance of the features.

### 4.1 Corpus collection

We implement the MLS-based RIR measurement on laptops. The built-in loudspeaker on laptop plays a MLS sequence and the microphone records the sound signal synchronously. The whole playing and recording process last about 18 seconds. Then a fast devolution algorithm is running on laptop to compute RIRs. In the end, a 2.8 second CSV file (16-bit 44kHz) recording the RIR is generated. We collected RIR samples in 10 different functional areas in Cory Hall locating at the campus of UC Berkeley, as listed in Table. 1. These 10 areas include both closed spaces and open spaces. "Closed" means that the area is surrounded by wall materials, such as offices and conference rooms. For "open" spaces, there is no clearly defined boundary, such as hallway. The kitchen investigated in our experiments is an area that connects two open aisles and thereby is an open space in our definition. These areas also vary in volumes, wall materials and furnishings. A description of their environments during data collection is given in Table. 1. All of the 10 areas are controlled by different lightings, which allows the lighting system to turn off individual zone to save energy. The experimenter set up the SoundLoc system at two positions in each area. For each position, 50 samples are collected. All the experiments are carried out in ordinary workdays. Hence, the majority of our samples are collected with random background noise, such as speaking, footsteps, door creaky noise and HVAC sounds.

### 4.2 Experiment design

#### 4.2.1 Experiment A

In this experiment, we aim at examining the distinctiveness of the features. We use the 1000 samples (10 areas × 2 positions × 50 samples) described in the last section to build our training sets and testing sets and conduct a 10 fold cross validation.

#### 4.2.2 Experiment B

This experiment aims at examining the ability of SoundLoc to deal with noise. In particular, we collected 100 RIR samples in the conference room during and after meetings. During the meeting, there exist successive talking and moving noise in the recordings and RIRs are computed from these noisy recordings. The testing set includes only noisy samples for conference room. The training set includes quiet samples for conference room and samples described in *Experiment A* for other places. A 5-fold cross validation is carried out in this experiment. Compared with *Experiment A*, this experiment is potentially challenging because the testing set is based on completely different samples that the model was trained on.

### 4.2.3 Experiment C

The purpose for this experiment is to test the time-stationarity of the features. We conduct the second visits in a different day to three areas: Office B, stairs and cubicle zone. These three locations are randomly picked from the places that are available. We do not exclude any samples after we see the result. For each place, 100 samples are collected. In the classification stage, the training set includes only the samples from the first visit while the test set includes only the samples from the second visit. 10 fold cross-validation is carried out.

## 5. RESULTS AND DISCUSSION

In this section, we evaluate the SoundLoc from the following aspects: distinctiveness, responsiveness, compactness, efficient-computability, noise-robustness and time-invariance. This is the DESENT criteria proposed in [10] to assess the performance of a fingerprint-based localization technique.

### 5.1 Parameter Study

As the reverberation dominates at the beginning of RIRs and is gradually overwhelmed by noise (Fig. 4), we truncate RIRs with a rectangular window in order to extract the reverberation information and diminish the influence of noise. Intuitively, more noise effect will be eliminated as the window length $t_{window}$ gets larger, but more information about reverberation will also be excluded in the meanwhile. And in practice, the following condition must be guaranteed for implementation of NAER,

$$t_{window} \geq \max\{RT_1,...,RT_n\} \quad (9)$$

where $n$ is the number of locations investigated. Fig. 8 provides insights for choosing an "optimal" $t_{window}$. As can be seen, in the region where (9) is satisfied, the accuracy increases when a smaller window is applied to RIRs. However, the accuracy drops dramatically once the window length is below RT. The minimum accuracy of labeling each room follows the same pattern. In our paper, $t_{window}$ is chosen to be 1.5 second, i.e. only the first half of RIRs are used for feature extraction.

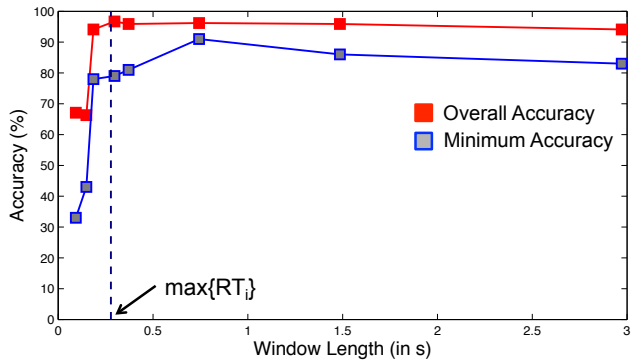

**Figure 8.** Overall accuracy and minimum accuracy increases as the window size decreases above RT. The accuracy drops significantly when the window size decreases blow RT.

### 5.2 Distinctiveness

In order to support localization, fingerprints should have a good separation between distinctive areas. In other words, there should be a one-to-one mapping from a room label to a feature distribution. The dissimilarity of distributions can be measured by Jensen-Shannon (JS) divergence, which is known for its unique capability of measuring a divergence between more than two probability distributions. JS divergence is given by the following formula:

$$JS(P_1,...,P_M) = \sum_{i=1}^{M} \pi_i KL(P_i \| \overline{P}) \quad (10)$$

where $\overline{P} = \sum_{i=1}^{M} \pi_i P_i$ is the mixed distribution, $\pi_i$ represents the weight for the distribution $P_i$, $\pi_i \in [0,1]$ and $\sum_{i=1}^{M} \pi_i = 1$. $KL(P_i \| \overline{P})$ is the Kullback-Leibler divergence, defined as

$$KL(P_i \| \overline{P}) = \sum_{x} \ln\left(\frac{P_i(x)}{\overline{P}(x)}\right) P_i(x) \quad (11)$$

JS divergence is a weighted sum of KL divergence and measures the distinctiveness of multiple distributions by considering how far each of the distributions deviates from the mixed distribution. The larger the JS divergence is, the better the separability the feature has achieved.

We verify the distinctiveness of a certain feature by classical permutation test. The idea is to randomly permute the labels of room labels and each time obtain a JS divergence. The null hypothesis is that the observed JS divergence for a given feature is independent of the room labeling, namely

$$H_0^{Feat.} : JS_{Observed}^{Feat.} = JS_{Permuted}^{Feat.} \quad (12)$$

where $H_0^{Feat.}$ denotes the null hypothesis for a certain feature. If the observed JS divergence significantly deviates from the mean of the JS divergence distribution in permutation test, we can reject the null hypothesis, i.e. the feature is distinctive for different locations.

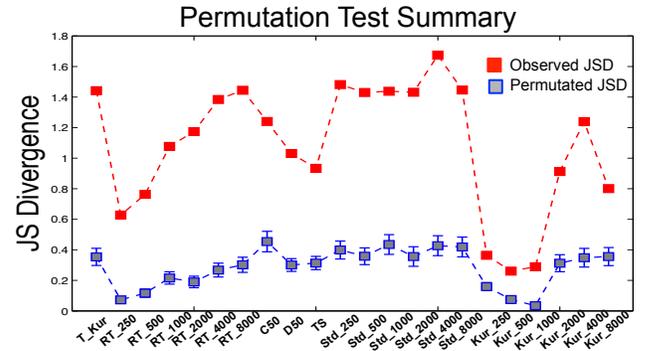

**Figure 9.** Permutation test summary. The observed JS divergence is significantly larger than the distribution of JS divergence obtained in permutation test and the null hypothesis should be rejected.

The result of the permutation test is presented in Fig. 9. The error bar specifies the quadruple standard deviation below and above the mean of JS divergence distribution in permutation test. As can be seen, the observed JS divergence is significantly larger than the distribution obtained in the permutation test. The result of p-value of significance testing is 2e-4, which shows that the probability of obtaining a JS divergence as extreme as observed under the null hypothesis is extremely small; therefore, the null hypothesis is rejected. We conclude that the features presented above can achieve high separability for different locations.

Within the context of localization, we further evaluate the feature distinctiveness by measuring the localization accuracy. The sequential floating forward selection (SFFS) algorithm was used to select a set of features to minimize the prediction error [24]. One essential reason for feature selection is to avoid overly complex models with respect to the number of features employed, since a sparse model is more robust to changes under different circumstances. For instance, we expect slight variations for feature distribution at different background noise levels, as discussed in Section 5.3, or at different times, as in Section 5.4. A robust set of features should be able to still distinguish the location from other in such conditions.

The sequential floating forward selection (SFFS) is an extension to the Plus-L minus-R selection (LRS) with flexible backtracking capabilities. In each iteration the number of features to be added and removed are determined from the data. Briefly, denote the feature set in iteration $i$ as $S^i$, the algorithm adds feature $f_k$ which satisfies:

$$f_k = \arg\min_f \left( (X,Y)_1^n ; S^{i-1}, f \right) \quad (13)$$

where $(X,Y)_1^n$ is the testing set with $n$ samples, $g(\cdot;\Theta)$ is the loss function when we use a feature set of $S^{i-1}, f$. The function is chosen as the misclassification rate aggregated for all the room labels. In the same iteration we also examine if by removing feature $f_v$ we can further reduce the misclassification error, and the removal process stops until we can no longer improve the classification result. Then the algorithm retries to add features in the optimal set in the next iteration. [24] provides a detailed algorithm for SFFS.

We use overall accuracy as the criterion for SFFS. Fig. 10 shows the classification confusion matrix for the room identification using optimal feature set listed in Table 3. The overall accuracy is 97.8%. When all features are used for classification, the overall accuracy is 95.9%.

|  | Office A | Office B | Office C | Office D | Conf. Rm | Lab | Kitchen | Hallway | Stairs | Cubicle |
|---|---|---|---|---|---|---|---|---|---|---|
| Office A | 98 |  |  | 2 |  | 1 |  |  |  |  |
| Office B | 1 | 97 |  | 4 |  |  |  |  |  |  |
| Office C |  |  | 100 | 0 |  |  |  |  |  |  |
| Office D |  | 3 |  | 93 |  |  | 1 |  |  |  |
| Conf. Rm |  |  |  |  | 100 |  |  |  |  |  |
| Lab | 1 |  | 1 |  |  | 99 | 2 |  | 2 |  |
| Kitchen |  |  |  |  |  |  | 94 |  |  |  |
| Hallway |  |  |  |  |  |  |  | 99 |  |  |
| Stairs |  |  |  |  |  | 1 | 3 |  | 98 |  |
| Cubicle |  |  |  |  |  |  |  |  |  | 100 |

**Figure 10.** Confusion matrix for *Experiment A*. The features used are from optimal set list in Table 3. The overall accuracy is 97.8%.

**Table 3. Optimal feature set created by SFFS**

| Temporal Feat. | Kur. |
|---|---|
| Spectral Feat. | Std. 1000 Hz, Std. 2000 Hz, Std. 4000 Hz, Std. 8000 Hz, Kur 1000 Hz, Kur. 2000Hz, Kur. 4000 Hz, Kur. 8000 Hz |
| Energetic Feat. | RT 1000 Hz, RT 2000 Hz, RT 4000 Hz, RT 8000 Hz, C50, D50, CT |

### 5.3 Noise-robustness

Generally speaking, any acoustic localization system exploits the location information hidden behind the recordings. Inevitably, there exist some transient changes that are independent of the position and carries no useful information for location estimation. For instance, speech noise leads to an over 20% accuracy drop in the localization method based on ambient background sound sensing [10]. Therefore, noise-robustness is a challenging issue in acoustic localization system.

Our localization technique leverages information from RIRs. They are computed using MLS excitation and cross-correlation technique. Since the phase spectrum of MLS is strongly erratic with a uniform density of probability in the $[-\pi,+\pi]$ interval, transient noise like clicks, footsteps etc. will be randomized and transformed into benign noise distributed evenly over the entire impulse response. Therefore, MLS-based RIR measurement should be expected to be immune to extraneous noise of all kinds theoretically. We design *Experiment B* to test the noise-robustness of SoundLoc. In this experiment, test and training set are noisy and quiet samples collected in the conference room, respectively. The accuracy of labeling the conference room is used as the indicator of noise-robustness. The result is presented in Table 4. The accuracy is very poor when all features or optimal features determined from *Experiment A* are used for classification. This result shows that the transient noise cannot fully weakened by using MLS technique in practice and some features in our feature pool are sensitive to noise. Again, we use SFFS and try to determine which feature is least sensitive to noise. The reselected features are listed in Table 5. They tend to exclude voice band, which is approximately 80-260 Hz. When using these selected features for classification, 98% accuracy can be achieved for labeling conference room.

**Table 4. Results summary of *Experiment B***

| Location | Accuracy | | |
|---|---|---|---|
|  | All Feat. | Exp. A Feat. | Reselected Feat. |
| Conf. Rm. | 15% | 24% | 98% |

**Table 5. Noise-robust feature set created by SFFS**

| Spectral Feat. | Std. 500 Hz, Std. 1000 Hz, Std. 2000 Hz, Std. 4000 Hz, Kur. 1000 Hz, Kur. 2000Hz |
|---|---|
| Energy Feat. | RT 500 Hz, RT 1000 Hz, RT 8000Hz, C50, D50, CT |

### 5.4 Time-invariance

A useful fingerprint should be relatively stationary over time. We design *Experiment C* to test the time-invariance of SoundLoc. In this experiment, test and training set are from separate visits. The result is summarized in Table 6. In general, using data from completely different visits for training results in a slightly lower accuracy than that when training and testing data come from the same visit. For Office B, the accuracy suffers from a dramatic fall when different visit samples are used for training. However, for stairs and cubicle, the accuracy remains above 99%. The reason for this is that stairs and cubicle are very different from other locations investigated in our paper, while the 4 offices in our experiment have similar geometry, wall materials and furnishings. The features do vary somewhat, which leads to confusion of very similar environments. We also test the features that are chosen by SFFS in *Experiment A*. The accuracy using this feature set is lower than that when all features are used for classification. That's because some features that are invariant in longer time scale but do not lead to the best accuracy are excluded during feature selection in *Experiment A*. We reselect the most time-stationary features using SFFS, as listed in Table 7. Higher than 93% accuracy has been achieved with the reselected features.

Table 6. Results summary of *Experiment C*

| Location | Training Type | Accuracy | | |
|---|---|---|---|---|
| | | All Feat. | Exp. A Feat. | Reselected Feat. |
| Office B | Different Visit | 79% | 76% | 95% |
| | Same Visit | 93% | 97% | 93% |
| Stairs | Different Visit | 99% | 98% | 99% |
| | Same Visit | 99% | 98% | 99% |
| Cubicle | Different Visit | 100% | 100% | 100% |
| | Same Visit | 99% | 100% | 99% |

Table 7. Time-invariant feature set created by SFFS

| Temporal Feat. | Kur. |
|---|---|
| Spectral Feat. | Std. 250 Hz, Std. 500 Hz, Std. 1000 Hz, Std. 2000 Hz, Std. 4000 Hz, Std. 8000 Hz, Kur. 250 Hz, Kur 500 Hz, Kur. 1000 Hz, Kur. 2000 Hz |
| Energy Feat. | RT 250 Hz, RT 500Hz, RT 2000 Hz, RT 4000 Hz, RT 8000 Hz, C50, D50, CT |

## 5.5 Comprehensive Feature Selection

The previous three experiments consider three experiments consider three different data collection and testing scenarios. Experiment A is when the distribution of testing and training data do not differ by noise or time. Experiment B and C consider the effect of noise and time transition on RIR respectively. We also conducted an additional experiment that considers all the settings by combining the noise- and time-corrupted data and use the mixed data for feature selection. In reality it is likely that the data is collected and tested under any one of the four scenarios.

Table 8 organizes the features by the number of time they are selected in each case. Features that are consistently selected, such as RT 8000 Hz, CT, and Std. 4000 Hz, are very likely to perform well in practice, since they stand the test of noise-robustness and time-invariance. Those who are selected more than two times are also good features since their inclusion can enhance the classification performance in most situations. We notice that some low frequency features are not likely to be selected, since it often suffers from human talking and ambient noises. As a guideline, the user is advised to include features sequentially from the first row to the last row in Table 8, depending on the complexity of model that is desired. Usually sparse model is good for generality and more complex model is good if the situation is not highly dynamic.

Table 8. Comprehensive feature selection table

| # time selected | Features |
|---|---|
| 4 out of 4 | RT 8000 Hz, CT, Std. 4000 Hz, Kur. 1000 Hz, Kur. 2000 Hz |
| 3 out of 4 | Time Kur., RT 2000 Hz, RT 4000 Hz, C50, D50, Std. 500 Hz, Std. 1000 Hz, Std 2000 Hz, Std 8000 Hz, Kur. 1000 Hz, Kur. 8000 Hz |
| 2 out of 4 | RT 250 Hz, RT 500 Hz, RT 1000 Hz, Kur 4000 Hz, Kur 8000 Hz |

## 5.6 Computation Effort

In this part, we evaluate the performance of SoundLoc based on its responsiveness and memory requirement.

The time taken for SoundLoc to identify the room mainly lies on three processing steps: RIR collection, RIR computation and feature extraction. The collecting and computing for one RIR in our corpus last 32 second, which accounts for the majority of SoundLoc's time cost. The feature extraction step is very effective, with most features computation requiring just an accumulative summation with complexity $\Theta(n_{RIR})$, where $n_{RIR}$ denotes the length of RIR. Feature computation time is mainly spent on spectral features that require a fast Fourier transform (FFT), giving a runtime complexity $\Theta(n_{RIR}\log(n_{RIR}))$. Another time cost comes from computing RT within different octave bands. Convolving RIR with octave filters has a complexity of $\Theta(r_{spec.}n_{oct.}n_{RIR})$, where $n_{oct.}$ denotes the length of octave filters and $r_{spec.}$ is resolution of octave bands. As $n_{oct.}$ and $r_{spec.}$ are negligible compared with $n_{RIR}$, the overall complexity of feature extraction will be $\Theta(n_{RIR}\log(n_{RIR}))$. Moreover, we can truncate RIR with a window in order to reduce the feature extraction time. According to Fig. 7, we can select a RIR length that can achieve relatively high accuracy and requires a moderate computation time in the meanwhile. The time taken to extract features for one RIR sample is 0.2 second in our experiment.

We also evaluate the features in terms of compactness, which describes the memory requirements of a fingerprint. The original RIR requires 1.5 Mega bytes of storage. However, only 188 bytes are needed to save the 22 features in total. We can further reduce the memory requirements by feature selection algorithm aforementioned.

## 5.7 Energy Footprint Optimization

In this part, we evaluate the SoundLoc based on its ability to lower the energy consumption whilst remain robust when faced with changing conditions. The overall energy footprint can be considered from two aspects: firstly, how much energy is required for collecting a single RIR sample; secondly, how many samples should be included in order to train a reliable model for a room.

### 5.7.1 Energy concerning single RIR collection

The power consumption of collecting a RIR sample involves the loudspeaker to emit a sound excitation and hardware to perform calculations including feature computation, digit-to-analog converting and vice versa. The relationship between acoustic power and sound power level is given by the following formula,

$$P = P_0 \cdot 10^{\frac{L_P}{10}} W \qquad (14)$$

For instance, 100 dB, which is almost the strongest sound level and begins to hurt human's hearing, corresponds 0.01 W. Therefore, the power for loudspeaker to play MLS is negligible compared with the power consumed by CPU and soundcard to process the digital signal. The following discussion will focus on the power consumption for hardware to process data.

The first parameter we consider here is sampling frequency. According to Shannon-Nyquist Sampling Theorem, if we want to reconstruct the analog signal, the sampling frequency is required to be at least twice the upper bound of frequency component of the original analog signal. That is, in order to inspect the room characteristics over higher frequency bands, the sampling frequency should be increased correspondingly. Fig. 11 shows that inclusion of higher frequency features leads to an increase of overall accuracy. The horizontal axis indicates the maximum octave bands center frequency involved in our feature pool. For most places investigated, the identification accuracy slightly changes as the frequency range inspected shrinks. However, some location, such as Office B, experiences a significant decrease of accuracy when we exclude the features of high frequency. Moreover, the sampling frequency is related to the time required for emitting excitation MLS signal by

$$t_{excitation} = \frac{N_{reps} \cdot L_{mls}}{f_s} \quad (15)$$

where $N_{reps}$ is the repetition times of MLS sequence, $L_{mls}$ is the length of MLS and $f_s$ denotes the sampling frequency. In addition to higher accuracy, increasing $f_s$ also benefits in terms of less RIR sample collection time. However, more power is needed for both sound card and CPU to process the increased size of audio as $f_s$ increases. And there is an upper bound for sampling frequency. $f_s \leq \frac{N_{reps} \cdot L_{mls}}{RT}$ should be satisfied in order to avoid time-aliasing.

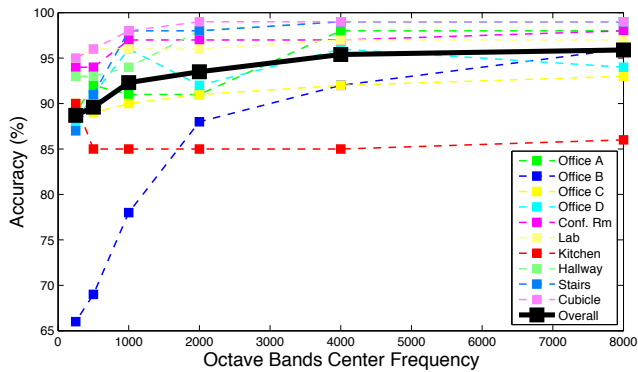

**Figure 11.** The effect of including features in higher frequency bands on the classification accuracy.

$N_{reps}$ is another parameter that we can tune. In practice, we employ repetitive MLS stimuli and average the received signal in order to reduce the influence of background noise. For every doubling the number of time responses averaged, neglecting quantization and coherent noise, the signal-to-noise ratio should theoretically increase by 3dB [25]. The tradeoff is measurement time, i.e. measurement power consumption, versus increased noise immunity. Similarly, it has been showed that the signal-to-noise ratio for the MLS sequence increases by 3dB when the length of the MLS sequence $L_{mls}$ is doubled [26]. However, measurement cost also comes in.

### 5.7.2 Energy concerning training sample size

Next, we will consider the overall energy footprint for modeling a specific location, i.e. the number of samples that are required for reliably labeling this place. There are several reasons to optimize the number of training sets for SoundLoc. Firstly, training labels are often costly and time-consuming to obtain. Secondly, it requires storage space on the mobile platform, so a large set of training samples might limit the number of places in the memory. Also, more training samples demand more computational power, which might represent a bottleneck on the battery-powered device. To study the effects of training size on classification accuracy, we vary the size of the training sets to train an array of popular classifiers and plot the results in Fig. 12.

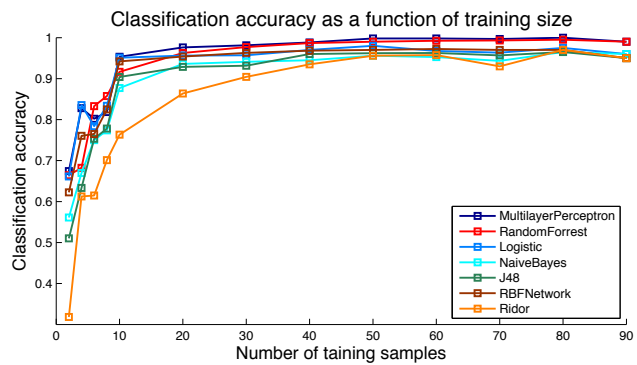

**Figure 12.** The effect of the number of training samples on the classification accuracy for various algorithms. All the models are implemented by the Weka machine learning toolkit [27]. The required number of training samples are selected randomly from the training sets. The rest of the samples are used as testing set.

As can be seen, the classification accuracy generally improves as the number of training samples increase. Most methods converge to an optimal classification rates when the number of training samples is from ten (10) to twenty (20). For the top algorithms such as Multilayer Perceptron and Random Forrest, the accuracy achieves 95.33% and 91.67% respective with only ten (10) samples. The satisfactory performance with only limited number of training samples is attributed to the separability, noise robustness, and time invariance of the sound features. Relaxing the requirement of training samples directly benefits the energy efficiency of SoundLoc, as well as easiness to implement.

## 6. CONCLUSION AND FUTURE WORK

We have presented SoundLoc, a room identification system exploiting the acoustic properties of the room. The acoustic properties are described quantitatively by various features extracted from room impulse response. We build a cheap MLS-based RIR measuring system using internal speakers and microphones on laptops. A noise adaptive reverberation extraction algorithm is developed to deal with feature extraction from the noisy RIRs. The algorithm is shown to be effective to extract reverberation time when the sound energy decay is dominated by direct sound and noise. Using this measurement system, we collect more than 1000 RIR samples in different locations, with different noise background and during separate days. The acoustic features we extracted are shown to be distinctive, robust and efficient to compute. 97.8% of overall accuracy has been achieved for 10 rooms' identification.

Moreover, the training sample size can be reduced to 10 samples while 95.3% accuracy can still be achieved.

For future work, we want to explore the application of acoustic features more than indoor localization. Automatic mapping is an emerging field in recent years. It aims at building a map with landmarks for an unmapped building. Right now, a robot equipped with sensors such as laser-based ranging and cameras are used for mapping. It is expensive and time-consuming to collect training samples. However, SoundLoc is cheap and energy efficient. A room can be identified in the map with very small dataset. It is potential and promising to be implemented for automatic mapping a building of interest organized by areas with distinctive acoustic properties.

# 7. ACKNOWLEDGMENTS

This research is funded by the Republic of Singapore's National Research Foundation through a grant to the Berkeley Education Alliance for Research in Singapore (BEARS) for the Singapore-Berkeley Building Efficiency and Sustainability in the Tropics (SinBerBEST) Program. BEARS has been established by the University of California, Berkeley as a center for intellectual excellence in research and education in Singapore.

# 8. REFERENCES

[1] Erickson, V. L., Achleitner, S., and Cerpa, A. E. POEM: Power-efficient occupancy-based energy management system. IEEE IPSN (2013).

[2] Zhen, Z., Jia, Q., Song, C. and Guan, X. An indoor localization algorithm for lighting control using RFID. IEEE ENERGY (2008).

[3] Williams, A., Ganesan, D., and Hanson, A. Aging in Place: fall detection and localization in a distributed smart camera network. In Proc. Intl. Conf. on Multimedia (2007), pp. 892-901.

[4] Want, R., Hopper, A., Falcao, V. and Gibbons, J. The active badge location system. ACM Transactions on Information Systems (Jan. 1992), 91-102.

[5] Haeberlen, A., Flannery, E., Ladd, A. M., Rudys, A., Wallach, D. S., and Kavraki, L. E. Practical robust localization over large-scale 802.11 wireless networks. In Proc. Intl. Conf. on Mobile Computing and Networking (MobiCom) (2004), pp. 70–84.

[6] Bahl, P. and Padmanabhan, V. N. RADAR: An in-building rf-based user location and tracking system. IEEE INFOCOM (2000)

[7] Youssef, M. and Agrawala, A. The Horus WLAN location determination system. In Proc. Intl. Conf. on Mobile Systems, Applications, and Services (MobiSys) (2005), pp. 205-218.

[8] Jiang, Y., Pan, X., Li, K., Lv, Q., Dick, R. P., Hannigan, M. and Shang, L. ARIEL: Automatic Wi-Fi based room fingerprinting for indoor localization. Intl. Conf. on Ubiquitous Computing (UbiComp) (2012), pp. 441-450.

[9] Azizyan, M., Constandache, I. and Choudhury, R. R. SurroundSense: mobile phone localization via ambience fingerprinting. In Proc. Intl. Conf. on Mobile Computing and Networking (MobiCom) (2009), pp. 261-272.

[10] Tarzia, S. P., Dinda, P. A., Dick, R. P. and Memik, G. Indoor localization without infrastructure using the acoustic background spectrum. Intl. Conf. on Mobile Systems, Applications, and Services (MobiSys) (2011), pp. 155-168.

[11] Lu, H., Pan, W., Lane, N. D., Choudhury, T., and Campbell, A. T. SoundSense: scalable sound sensing for people-centric applications on mobile phones. In Proc. Intl. Conf. on Mobile Systems, Applications, and Services (MobiSys) (2009), 165–178.

[12] Constandache, I., Agarwal, S., Tashev, I. and Choudhury, R. R. Daredevil: indoor location using sound. ACM SIGMOBILE Mobile Computing and Communications Review (Apr. 2014), pp. 9-19.

[13] Peters, N., Lei, H. and Friedland, G. Name that room: room identification using acoustic features in a recording. In Proc. Intl. Conf. on Multimedia (2012), pp. 841-844.

[14] Shabtai, N. R., Zigel, Y., Rafaely, B. Estimating the room volume from room impulse response via hypothesis verification approach. IEEE SSP (2009).

[15] G. Stan, J. Embrechts, and D. Archambeau. Comparison of different impulse response measurement techniques. JAES, 50, 4 (Apr. 2002), 249-262.

[16] Oppenheim, A. V. and Schafer, R. W. Discrete-time Signal Processing.

[17] Rife, D. D. and Vanderkooy, J. Transfer-function measurement with maximum-length sequences. JAES, 37, 6 (June 1989), 419-444.

[18] Borish, J. and Angell, J. . An efficient algorithm for measuring the impulse response using pseudorandom noise. JAES, 31, 7/8 (Aug. 1983), 478-488

[19] Jetzt, J. J. Critical distance measurement of rooms from the sound energy spectral response. J. Acoust. Soc. Am., 65, 1204 (1979)

[20] Moore, A. H., Brookes, M. and Naylor, P. A. Roomprints for forensic audio applications. IEEE WASPAA (2013)

[21] Kuttruff, H. Room Acoustics. London: Taylor & Francis (2000)

[22] Guski, M. and Vorlander, M. Noise compensation methods for room acoustical parameter evaluation. In Proc. Intl. Symposium on Room Acoustics (Jun. 2013).

[23] Chu, W. T. Comparison of reverberation measurements using Schroeder's impulse method and decay curve averaging method. J. Acoust. Soc. Am., 63, 1444 (1978).

[24] Somol, P., Pudil, P., Novovičová, J., & Paclík, P. (1999). Adaptive floating search methods in feature selection. Pattern recognition letters, 20, 11/13 (Nov. 1999), 1157-1163.

[25] Fincham and Laurie, R. Refinements in the impulse testing of loudspeakers. JAES, 74 (Oct. 1983)

[26] Bleakley, C. and Scaife, R. New formulas for predicting the accuracy of acoustical measurements made in noisy environments using the averaged m-sequence correlation technique. J. Acoust. Soc. Am., 97, 1329 (1995)

[27] Hall, M., Frank, E., Holmes, G., Pfahringer, B., Reutemann, P. and Witten, I. H. The WEKA Data Mining Software: An Update; SIGKDD Explorations, 11, 1 (2009).